# A connection between star formation activity and cosmic rays in the starburst galaxy M 82

V. A. Acciari [1,23], E. Aliu [2], T. Arlen [3], T. Aune [4], M. Bautista [5], M. Beilicke [6], W. Benbow [1], D. Boltuch [2], S. M. Bradbury [7], J. H. Buckley [6], V. Bugaev [6], K. Byrum [8], A. Cannon [9], O. Celik [3], A. Cesarini [10], Y. C. Chow [3], L. Ciupik [11], P. Cogan [5], P. Colin [12], W. Cui [13], R. Dickherber [6], C. Duke [14], S. J. Fegan [3], J. P. Finley [13], G. Finnegan [12], P. Fortin [15], L. Fortson [11], A. Furniss [4], N. Galante [1], D. Gall [13], K. Gibbs [1], G. H. Gillanders [10], S. Godambe [12], J. Grube [9], R. Guenette [5], G. Gyuk [11], D. Hanna [5], J. Holder [2], D. Horan [16], C. M. Hui [12], T. B. Humensky [17], A. Imran [18], P. Kaaret [19], N. Karlsson [11], M. Kertzman [20], D. Kieda [12], J. Kildea [1], A. Konopelko [21], H. Krawczynski [6], F. Krennrich [18], M. J. Lang [10], S. LeBohec [12], G. Maier [5], S. McArthur [6], A. McCann [5], M. McCutcheon [5], J. Millis [22], P. Moriarty [23], R. Mukherjee [15], T. Nagai [18], R. A. Ong [3], A. N. Otte [4], D. Pandel [19], J. S. Perkins [1], F. Pizlo [13], M. Pohl [18], J. Quinn [9], K. Ragan [5], L. C. Reyes [24], P. T. Reynolds [25], E. Roache [1], H. J. Rose [7], M. Schroedter [18], G. H. Sembroski [13], A. W. Smith [8], D. Steele [11], S. P. Swordy [17], M. Theiling [1], S. Thibadeau [6], A. Varlotta [13], V. V. Vassiliev [3], S. Vincent [12], R. G. Wagner [8], S. P. Wakely [17], J. E. Ward [9], T. C. Weekes [1], A. Weinstein [3], T. Weisgarber [17], D. A. Williams [4], S. Wissel [17], M. Wood [3], B. Zitzer [13]

[1] Fred Lawrence Whipple Observatory, Harvard-Smithsonian Center for Astrophysics, Amado, AZ 85645, USA. [2] Department of Physics and Astronomy and the Bartol Research Institute, University of Delaware, Newark, DE 19716, USA. [3] Department of Physics and Astronomy, University of California, Los Angeles, CA 90095, USA. [4] Santa Cruz Institute for Particle Physics and Department of Physics, University of California, Santa Cruz, CA 95064, USA. [5] Physics Department, McGill University, Montreal, QC H3A 2T8, Canada. [6] Department of Physics, Washington University, St. Louis, MO 63130, USA. [7] School of Physics and Astronomy, University of Leeds, Leeds, LS2 9JT, UK. [8] Argonne National Laboratory, 9700 S. Cass Avenue, Argonne, IL 60439, USA. [9] School of Physics, University College Dublin, Belfield, Dublin 4, Ireland. [10] School of Physics, National University of Ireland, Galway, Ireland. [11] Astronomy Department, Adler Planetarium and Astronomy Museum, Chicago, IL 60605, USA. [12] Department of Physics and Astronomy, University of Utah, Salt Lake City, UT 84112, USA. [13] Department of Physics, Purdue University, West Lafayette, IN 47907, USA. [14] Department of Physics, Grinnell College, Grinnell, IA 50112-1690, USA. [15] Department of Physics and Astronomy, Barnard College, Columbia University, NY 10027, USA. [16] Laboratoire Leprince-Ringuet, Ecole Polytechnique, CNRS/IN2P3, F-91128 Palaiseau, France. [17] Enrico Fermi Institute, University of Chicago, Chicago, IL 60637, USA. [18] Department of Physics and Astronomy, Iowa State University, Ames, IA 50011, USA. [19] Department of Physics and Astronomy, University of Iowa, Van Allen Hall, Iowa City, IA 52242, USA. [20] Department of Physics and Astronomy, DePauw University, Greencastle, IN 46135-0037, USA. [21] Department of Physics, Pittsburg State University, 1701 South Broadway, Pittsburg, KS 66762, USA. [22] Department of Physics, Anderson University, 1100 East 5th Street, Anderson, IN 46012. [23] Department of Life and Physical Sciences, Galway-Mayo Institute of Technology, Dublin Road, Galway, Ireland. [24] Kavli Institute for Cosmological Physics, University of Chicago, Chicago, IL 60637, USA. [25] Department of Applied Physics and Instrumentation, Cork Institute of Technology, Bishopstown, Cork, Ireland

**Although Galactic cosmic rays (protons and nuclei) are widely believed to be dominantly accelerated by the winds and supernovae of massive stars, definitive evidence of this origin remains elusive nearly a century after their discovery [1]. The active regions of starburst galaxies have exceptionally high rates of star formation, and their large size, more than 50 times the diameter of similar Galactic regions, uniquely enables reliable calorimetric measurements of their potentially high cosmic-ray density [2]. The cosmic rays produced in the formation, life, and death of their massive stars are expected to eventually produce diffuse gamma-ray emission via their interactions with interstellar gas and radiation. M 82, the prototype small starburst galaxy, is predicted to be the brightest starburst galaxy in gamma rays [3, 4]. Here we report the detection of >700 GeV gamma rays from M 82. From these data we determine a cosmic-ray density of 250 eV cm$^{-3}$ in the starburst core of M 82, or about 500 times the average Galactic density. This result strongly supports that cosmic-ray acceleration is tied to star formation activity, and that supernovae and massive-star winds are the dominant accelerators.**

M 82 is a bright galaxy located approximately 12 million light years from Earth, in the direction of the Ursa Major constellation [5]. For hundreds of millions of years, M 82 has been gravitationally interacting with nearby galaxies, including the larger spiral galaxy M 81 [6]. Over time, interactions with these neighbours have deformed M 82, creating an active starburst region in its centre with a diameter of ~1000 light years [7]. The Hubble Space Telescope reveals hundreds of young massive ($10^4$ to $10^6$ solar mass) clusters in this starburst region [8]. Throughout this compact region stars are being formed at a rate approximately 10 times faster than in entire "normal" galaxies like the Milky Way, and the supernovae rate is 0.1 to 0.3 per year [9, 10]. The intense radio-synchrotron emission observed in the central region of M 82 suggests a very high cosmic-ray energy density, about two orders of magnitude higher than in the Milky Way [11]. The region also contains a high mean (molecular) gas density of about 150 particles per cm$^3$, or about $10^9$ solar masses in total [12]. Given the high cosmic-ray and gas densities, M 82 has long been viewed as a promising target for gamma-ray observatories [7]. However, it was not detected above 100 MeV by the EGRET experiment [13], nor during previous very high energy (VHE, E >100 GeV) gamma-ray observations of M 82 with the Whipple 10-m [14] and HEGRA [15] experiments. The latter two set upper limits at ~10% of the flux from the Crab Nebula, the brightest steady VHE source in the sky. These limits are well above the sensitivity of the Very Energetic Radiation Imaging Telescope Array System (VERITAS).

VERITAS [16] is located in southern Arizona and has been fully operational since September 2007. It consists of a stereoscopic array of four 12-m diameter optical telescopes equipped with sensitive cameras (3.5° field-of-view) that detect short (few nanosecond) flashes of ultraviolet and blue light, known as Cherenkov radiation. This light is emitted in the electromagnetic cascade of secondary particles resulting from the interaction of a VHE gamma ray in the upper atmosphere. VERITAS has an energy threshold of ~100 GeV, an energy resolution of ~15%, and an angular resolution of ~0.1° per event.

M 82 was observed with VERITAS for a total of ~137 hours of quality-selected live time between January 2008 and April 2009 at a mean zenith angle of 39°. This exceptionally long exposure was taken entirely during periods of astronomical darkness and clear atmospheric conditions. The analysis of these data was performed with the standard VERITAS analysis procedure [17] using event-selection criteria optimised *a priori* for low-flux, hard-spectrum sources. An excess of 91 gamma-ray-like events (~0.7 photons per hour) above the estimated background (267 events) is observed from the direction of M 82 (see the Supplementary Information for more details). This excess corresponds to a post-trials statistical significance of 4.8 standard deviations (σ), or a chance probability of $7.7 \times 10^{-7}$, and represents the discovery of VHE gamma-ray emission from M 82 (see Figure 1). The observed differential VHE gamma-ray spectrum (see Figure 2) is best fitted using a power-law function with a photon index $\Gamma = 2.5 \pm 0.6_{stat} \pm 0.2_{syst}$. The measured gamma-ray flux is $(3.7 \pm 0.8_{stat} \pm 0.7_{syst}) \times 10^{-13}$ cm$^{-2}$ s$^{-1}$ above the 700 GeV energy threshold of the analysis, and no flux variations are observed. The luminosity of M 82 above 700 GeV inferred from the gamma-ray flux is $2 \times 10^{32}$ W, or about $2 \times 10^{6}$ times smaller than its far infrared (100 μm) luminosity [18].

At a flux of 0.9% of that observed from the Crab Nebula, M 82 is among the weakest VHE sources ever detected. Although VERITAS has detected several confirmed VHE sources near this flux, a large number of tests were performed to ensure systematic effects could not potentially create a spurious signal in the data (see the Supplementary Information). None of these tests give any indication that the observed signal is an artifact.

Prior to the VERITAS discovery of VHE gamma-ray emission from M 82, all known extragalactic VHE sources were clearly associated with an active galactic nucleus (AGN), an object powered by accretion onto a supermassive black hole. Although M 82 may host a supermassive black hole at its centre, it exhibits at most only a weak level of AGN activity [19]. On the other hand, the high rate of star formation in M 82 implies the presence of numerous strong shock waves in supernova remnants and around massive young stars. In the Milky Way, similar shock waves are known to accelerate electrons to very high energies, and they are suspected to likewise accelerate ions. This acceleration is expected to supply the cosmic rays which permeate both the Galaxy and M 82, and which produce diffuse gamma ray emission.

The most recent theoretical models [2, 3, 4, 7] predict a VHE gamma-ray flux from M 82 based on the acceleration and propagation of cosmic rays in the starburst core. These calculations are all close to the value measured by VERITAS. Using the model [3] shown in Figure 2, the cosmic-ray density in the starburst core of M 82 is estimated from the VHE flux to be ~250 eV cm$^{-3}$, or approximately 500 times the average Milky Way density. Although the cosmic-ray density of the M 82 core is significantly higher, the total cosmic-ray energy content of the two systems is similar since the volume of the Milky Way is about 500 times larger. The lifetime of cosmic-ray particles in the M 82 core is constrained to approximately one million years on account of energy losses though adiabatic cooling in the starburst wind and through collisions with interstellar gas nuclei. This is about 30 times shorter than the lifetime of the GeV-band particles in the Milky Way, which dominate the local cosmic-ray density. Thus a correspondingly larger source power is needed to replenish these particles in M 82 to maintain a similar cosmic-ray energy content. Interestingly, the estimated supernova rate in M 82 is about a factor of 30 larger than in the Milky Way. Thus, the VERITAS data show an enhancement in the cosmic-ray acceleration that matches the enhancement in energy input by massive stars and supernovae. This correlation strongly supports the long-held theory that these objects play a dominant role in cosmic-ray production.

Although the VERITAS data strongly indicate smaller shocks (e.g., those in supernova remnants) are the predominant cosmic-ray acceleration sites, it also cannot be ruled out that this acceleration occurs on larger (>30 light year) scales in a more distributed fashion [1]. Significantly lower estimates of the M 82 supernova rate [4] would also suggest other potential sources of cosmic-ray acceleration. However, alternative sources of mechanical energy for cosmic-ray acceleration, such as galactic rotation [1], can be ruled out.

The aforementioned theoretical models include significant contributions from both leptonic (e.g., electrons) and hadronic (e.g., ions) particle interactions, which are expected to give different VHE gamma-ray spectra (see Figure 2). Cosmic-ray ions create VHE gamma rays through collisions with interstellar matter. This process creates unstable particles called pi-mesons (pions). Electrically neutral pions directly decay into gamma rays. Charged pions eventually decay into neutrinos and electrons. The latter emit synchrotron radiation in the radio and infrared bands through interactions with the ambient magnetic field. The radio emission from these secondary electrons can be used to place an upper limit on the gamma-ray flux produced by cosmic-ray ions, thus helping to further discriminate between VHE gamma rays emitted by cosmic-ray ions and those coming from cosmic-ray electrons. The radio flux observed at 32 GHz frequency [20] implies that cosmic-ray ions would not produce a gamma-ray flux at 20 GeV higher than about $2.5 \times 10^{-9}$ cm$^{-2}$ s$^{-1}$, unless the magnetic field in M 82 is considerably weaker than the conventional estimate of 8 nT. An extrapolation of the VHE gamma-ray spectrum measured with VERITAS using the fitted power-law index $\Gamma$ = 2.5 would exceed that limit by a factor of two, whereas an extrapolation with $\Gamma$ = 2.3, within the uncertainty range of the fit, would satisfy the limit. The comparison suggests that either the true gamma-ray spectrum between 10 GeV and 1 TeV is slightly harder than our best-fit spectrum suggests, or the gamma-ray emission does not come predominantly from cosmic-ray ions.

The observed radio emission may also come from the relativistic cosmic-ray electrons accelerated in M 82. All electrons interact with ambient infrared photons, boosting them into the hard X-ray/soft gamma-ray band via inverse-Compton scattering. This non-thermal process contributes ~25% to the diffuse X-ray flux [21], the remainder of which originates from thermal emission of hot gas. Observational limits on the steady, non-thermal diffuse X-ray emission, place its luminosity at 5 keV photon energy not significantly higher than the VHE gamma-ray luminosity observed with VERITAS. These X-ray data provide a lower limit to the amplitude of the interstellar magnetic field at about a third of the current estimate ($B$ = 8 nT), and hence an upper limit to the absolute number of relativistic cosmic-ray electrons with about 1 GeV kinetic energy in M 82. Electrons of much higher kinetic energy (about 10 TeV) are needed to produce VHE gamma rays through inverse-Compton scattering of ambient infrared photons. Both theoretical considerations and a comparison of the observed VHE gamma-ray flux with limits on the cosmic-ray-induced X-ray flux suggest that the inverse-Compton emission should have a hard spectrum with power-law index around 2 between 100 keV and 100 GeV. Since >100 GeV electrons quickly lose their energy by inverse-Compton scattering and synchrotron emission, eventually preventing their further acceleration above a characteristic energy, the inverse-Compton radiation spectrum should also steepen and eventually show a cut-off. The identification of a cut-off in this spectrum, potentially observable by combining data taken with VERITAS and NASA's Fermi Gamma-ray Space Telescope, could demonstrate which type of cosmic-ray particle is responsible for the VHE emission.

The VERITAS measurements of M 82 also have implications for the interpretation of the striking correlation observed between the far-infrared (FIR; from warm dust) and the radio emission (from synchrotron radiation of cosmic-ray electrons) in starburst galaxies [22, 23]. Massive star formation is generally accepted as the origin of both [24], but consensus is lacking on how such a tight correlation is produced [25, 26, 27]. The VHE flux measured from M 82 places these models on a sound footing by providing an independent estimate of the cosmic-ray density. The observed VHE flux also requires a a hard cosmic-ray spectrum, $E^{-p}$ with $p$ near 2.1 to 2.3 [3, 4], placing new constraints on models of the radio-FIR correlation.

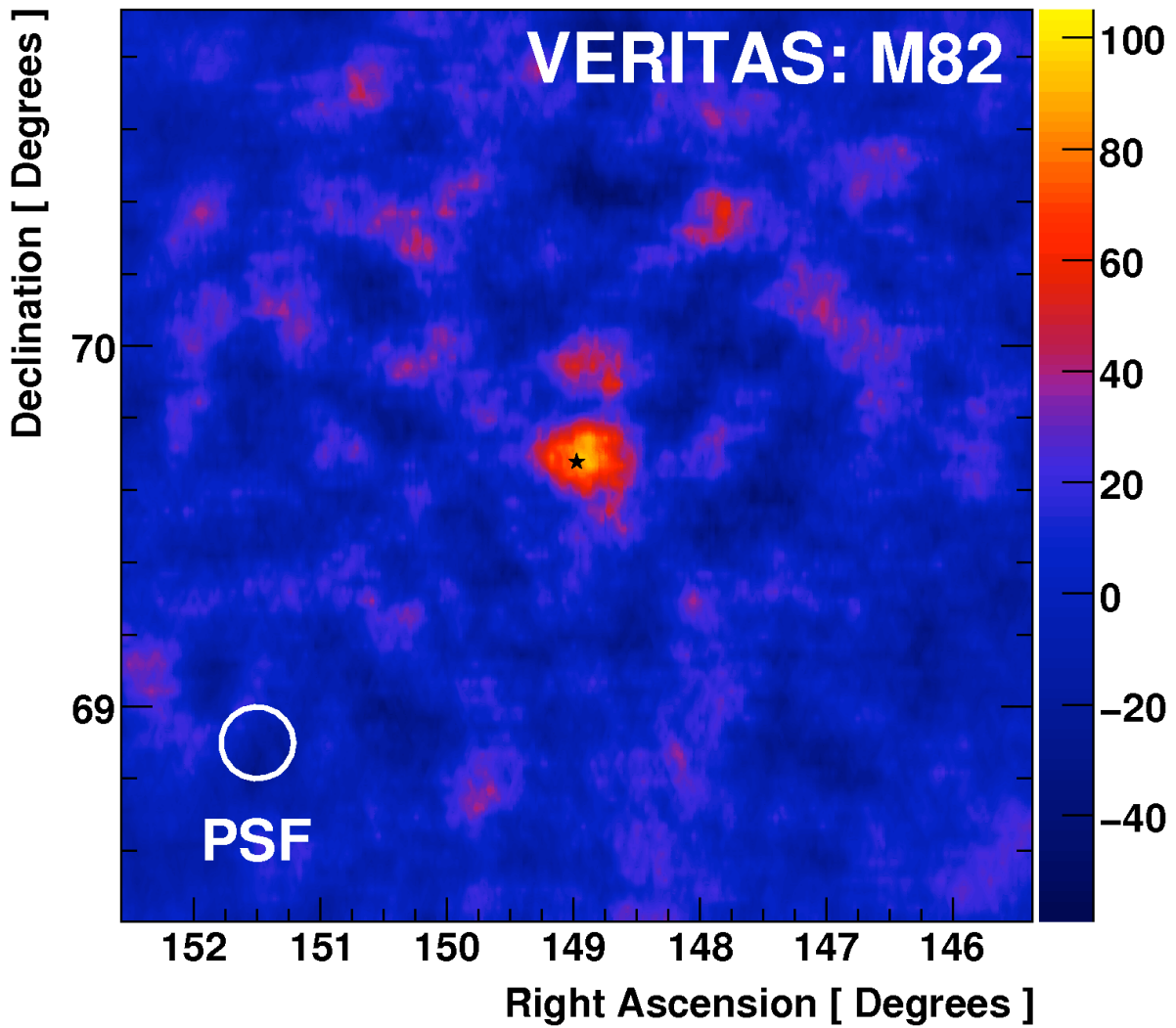

**Figure 1 Caption:** VHE image of the M 82 region. The sky map shows the measured excess (colour-scale) of gamma-ray-like events above the estimated background from a region centred on M 82. Each pixel contains the excess in a circular region of radius 0.1°. The map is over-sampled, and neighbouring pixels are thus correlated. The background for each point is estimated using an annulus centred on its position (the ring method [28]). The spatial distribution of the observed excess is consistent with that expected from a point-like source located near the core of M 82. The white circle represents the VERITAS point-spread function (68% containment) for individual gamma rays. The uncertainty in the source localisation is much smaller. The black star denotes the location of the core of M 82. The coordinates are for the J2000 epoch.

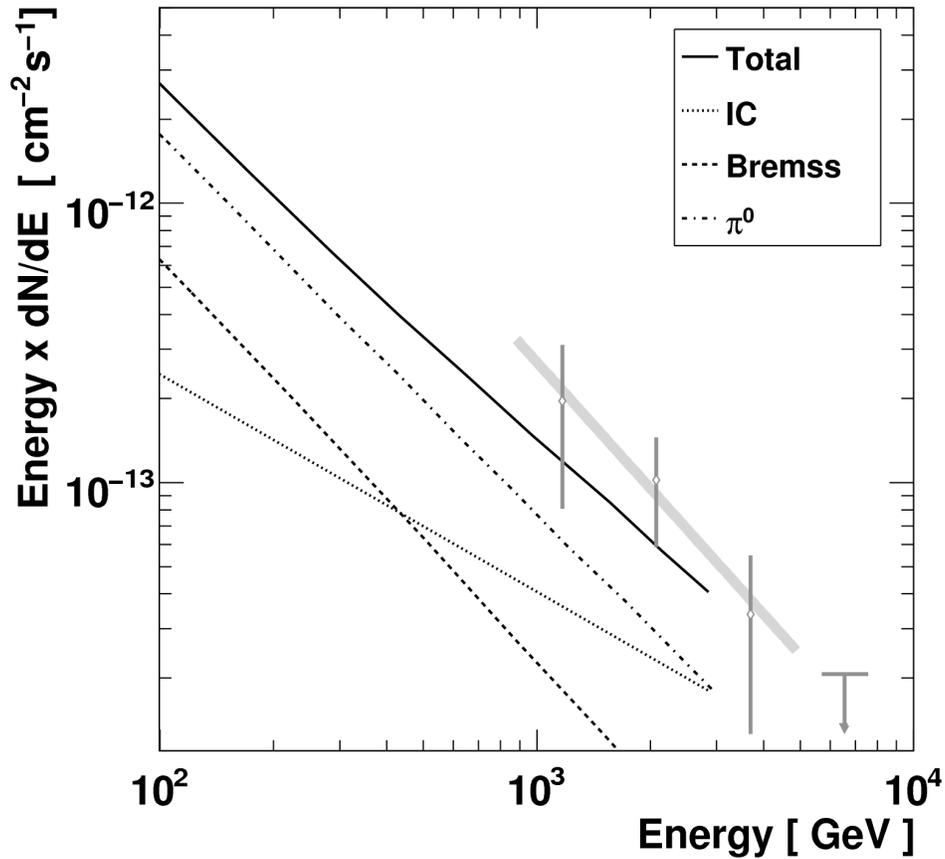

**Figure 2 Caption:** Gamma-ray flux compared to a theoretical prediction. The differential energy spectrum (E x dN/dE) of M 82 observed by VERITAS between ~0.9 TeV and ~5 TeV is shown. The data are given by open diamonds with 1σ statistical error bars, and can be fitted ($\chi^2$ = 0.1 with 1 degree of freedom) by a power-law function (thick gray line), dN/dE ~ (E / TeV)$^{-\Gamma}$, with $\Gamma$ = 2.5 ± 0.6$_{stat}$ ± 0.2$_{syst}$. The VERITAS flux upper limit (99% confidence level [29]) shown at ~6.6 TeV is above the extrapolation of the fitted power-law function at these energies. The thin lines represent a recent model [3] for the gamma-ray emission from M 82. The thin solid line is the total emission predicted, and the dashed lines represent components of this emission from the interactions of cosmic-ray ions with interstellar matter (π⁰ decay), and from radiation from cosmic-ray electrons through inverse-Compton scattering (IC) and Bremsstrahlung (Bremss). The IC and the π⁰ decay components are the dominant contributions of cosmic-ray electrons and ions, respectively. The markedly different spectral slopes of these dominant components should be noted.

**Supplementary Information** is linked to the online version of the paper at www.nature.com/nature.

**Acknowledgments**

This research is supported by grants from the US Department of Energy, the US National Science Foundation, and the Smithsonian Institution, by NSERC in Canada, by Science Foundation Ireland, and by STFC in the UK. We acknowledge the excellent work of the technical support staff at the FLWO and the collaborating institutions in the construction and operation of the instrument.

**Author Contributions**

VERITAS is a collaboration of scientists who jointly participate in all aspects of the scientific efforts: the taking of data, the development of software for analysis and simulations, the analysis of data and in the interpretation of the results. Every author has read the paper and agrees with the results.

**Author Information**

Correspondence and requests for materials should be addressed to Dr. Wystan Benbow (wbenbow@cfa.harvard.edu).

**Supplementary Information**

In our paper we present a detection of very high energy (VHE) γ-ray emission from the starburst galaxy Messier 82. Here we discuss the analysis methods used to extract the VHE gamma-ray signal, and summarize the systematic tests performed to ensure the signal is not an experimental artifact.

**1. Analysis Details**

A key aspect of detecting VHE gamma-ray emission using atmospheric-Cherenkov telescopes is eliminating as much as possible of the dominant background, which comes from the interaction of charged cosmic rays in the upper atmosphere. In this analysis, event-selection cuts on a number of parameters, derived from the images recorded by the VERITAS cameras, are used to reduce the number of background events [30, 31, 17]. For each particle cascade triggering VERITAS [32], the corresponding Cherenkov light illuminating the 3.5º field of view of each VERITAS camera [33] is recorded using 499 fast photo-multiplier tubes (pixels) that subtend ~0.15º each. The light measured in these pixels is digitised in 2 nanosecond samples by an FADC, and the digitised pixel pulses are calibrated and converted into an integrated charge proportional to the amount of light recorded [34, 35]. After the Cherenkov images are calibrated, each image is cleaned to remove extraneous pixels that have been triggered by night sky background light [17]. In order for any recorded Cherenkov image to be used in the analysis, it must have a minimum integral size of 1000 digital counts (~200 photoelectrons), must have an image centroid within 1.43° of the centre of the camera, and must contain at least five pixels surviving the standard image-cleaning procedure. An event is only considered if at least two of its camera images pass these pre-selection criteria. In addition, an event is rejected if there are surviving images in only the two closest (separation of 35 m, c.f. ~80 m for the other pairs) VERITAS telescopes. Other cuts are applied on the mean-scaled width (0.05 < *MSW* < 1.10) and length parameters (0.05 < *MSL* < 1.30) [17], as well as on the square of the separation of reconstructed arrival direction of the event from the putative source position ($\theta^2$ < 0.01 square degrees). As the VHE gamma-ray spectrum of M 82 is expected to be hard, the size cut was chosen *a priori* to be the same as that used for the analysis of Galactic sources observed by VERITAS, and such cuts are typically used by other VHE experiments for their analysis of weak, hard-spectrum sources. The remaining cuts were optimised using Crab Nebula data taken at similar (large) zenith angles to the M 82 observations. It is important to note that the ~1' angular diameter of M 82 is effectively point-like for VERITAS. As a result of the high cut on minimum image size and the relatively large zenith angle of these observations, the analysis threshold (700 GeV) for these data is considerably higher, and the sensitivity is somewhat less (due to the zenith angle), than for typical VERITAS observations.

Using the aforementioned selection criteria, a total of 358 gamma-ray-like events are counted in a circular region (radius 0.1º) centred on the direction of M 82. The background expected in this region (267 events) is estimated by counting the number of events (2941) in 11 identical regions in the same field of view. It should be noted that the location of M 82 was offset by 0.5º from the centre of the camera field of view, alternating in each of the four cardinal directions, during the observations. This enables the use of a reflected-region method [28], where each background region has the same magnitude of camera offset as M 82. The observed excess of 91 events has a statistical significance of 5.0 standard deviations ($\sigma$) following Equation 17 of [36]. Although weak-source, hard-spectrum selection cuts are commonly used in the literature and are well motivated for the observations of M 82, these cuts are different (larger size and smaller $\Theta^2$) from the standard cuts used in a typical VERITAS point-source analysis. In addition, a set of cuts most-sensitive to soft spectrum sources were also tried. To account for the different sets of cuts, the inclusion of a trials factor is appropriate for assessing the true statistical significance of this result. A conservative estimate of the total number trials (3) marginally decreases the post-trials significance to 4.8$\sigma$ (7.7 x $10^{-7}$ chance probability). Supplemental Figure 1 shows the radial distribution of the observed excess, which is consistent with that expected from a point-like source. The estimated background in Supplemental Figure 1 is also approximately flat, as expected. Supplemental Figure 2 shows the observed flux binned by monthly observation period. A fit of a constant flux to this light curve yields a $\chi^2$ of 11.5 for 9 degrees of freedom ($P(\chi^2) = 0.24$). Although the observed flux is consistent with being constant in time, flux variations of a factor of a few cannot be ruled out due to low statistics.

## 2. Systematic Tests

All data used in the analysis have passed rigorous checks to ensure the hardware is operating normally during the observations. The VERITAS signal from M 82, as well as the associated scientific quantities, are verified by several independent calibration and analysis chains. These packages each use different software, independently derived event-selection criteria, different methods for data-quality selection, and different instrument simulations. The reported significance is close to, but not, the highest value, measured in the independent analyses. The observed signal persists (5.1σ) using an alternate ring method [28] for estimating the background in the signal region, as well as when using a binned maximum-likelihood method for estimating the signal strength (i.e., not simply integrating the number of events in a circular region). The distribution of significance (excluding the observed signal) across the field of view centred on M 82 is well-described by a Gaussian function with mean zero and standard deviation one, as expected. The observed excess and significance increase with the observation time in a manner consistent with that of a steady signal, which is significant because VHE flux variability from M 82 is not expected. The observed excess also persists when each of the four telescopes is individually excluded from four otherwise identical analyses. To ensure that the optical light (magnitude 9.3) of M 82 does not contribute to the observed flux, the significances observed at the location of two comparably bright stars (magnitudes ~8.8 and ~9.0) in the same field of view were determined and found to be consistent with zero.

Although VERITAS has detected several confirmed VHE sources that are about two times brighter than M 82, it is the weakest source detected by the instrument so far. Therefore it is important to test whether an unknown systematic bias in VERITAS data might slowly create a false excess in the unusually long (~137 h live-time) data set presented here. A test data set with comparable exposure (~120 h live-time) was generated by combining the VERITAS data from several radio galaxies, galaxy clusters, dwarf galaxies and globular clusters. While some of these objects might emit VHE gamma rays, an identical analysis of these data results in a combined deficit of 4 events

(statistical significance of -0.2σ) compared to the background at the putative source locations. All of these systematic tests demonstrate that the gamma-ray signal observed from M 82 is genuine.

**Additional References:**

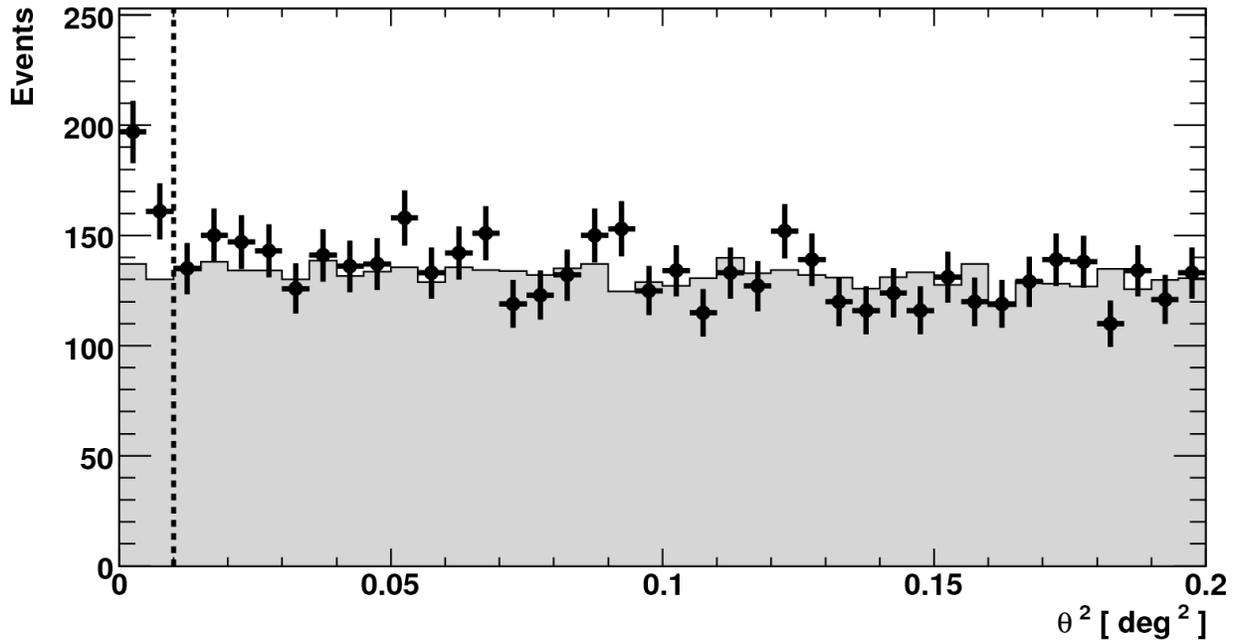

**Supplemental Figure 1:** The distribution of the square of the difference between the reconstructed arrival direction of an event and the direction of M 82. The points represent the measurement and the shaded region represents the estimated background. The vertical line shows the size of the circular integration region used in this analysis. The statistical errors correspond to one standard deviation ($\sigma$).

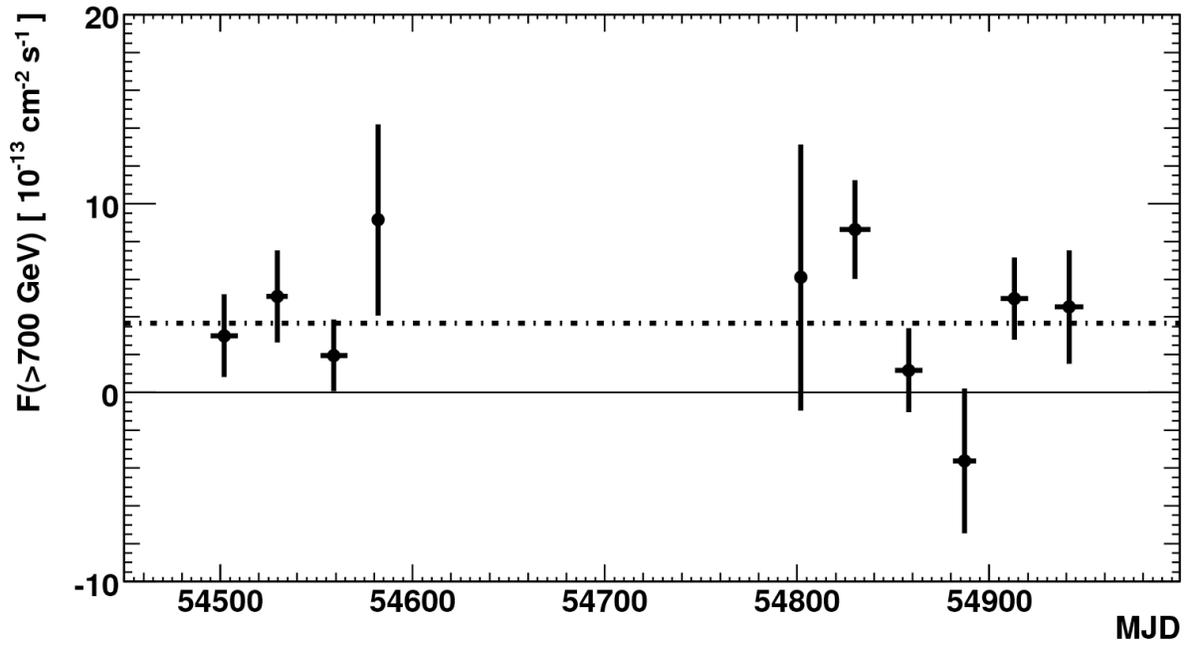

**Supplemental Figure 2:** The time-averaged flux observed from M 82 during each monthly observation period. The integral flux above 700 GeV is calculated assuming a photon index of $\Gamma = 2.5$. The fit of a constant value to the monthly flux is shown as a horizontal dashed line. Only the statistical errors (1σ) are shown. The horizontal bars on each point reflect the actual range of dates for each monthly exposure.